Stresa, Italy, 25-27 April 2007# A TWO-STEP ETCHING METHOD TO FABRICATE NANOPORES IN SILICON

*Gou-Jen Wang[1*], Wei-Zheng Chen[1], Kang J. Chang[2]*

[1]Department of Mechanical Engineering,
National Chung-Hsing University,
Taichung, Taiwan
Email:gjwang@dragon.nchu.edu.tw

[2]Department of Mfg. Systems Eng.
California State University,
Northridge, California, USA**ABSTRACT**

A cost effectively method to fabricate nanopores in silicon by only using the conventional wet-etching technique is developed in this research. The main concept of the proposed method is a two-step etching process, including a premier double-sided wet etching and a succeeding track-etching. A special fixture is designed to hold the pre-etched silicon wafer inside it such that the track-etching can be effectively carried out. An electrochemical system is employed to detect and record the ion diffusion current once the pre-etched cavities are etched into a through nanopore. Experimental results indicate that the proposed method can cost effectively fabricate nanopores in silicon.

*Keywords:* Nanopore fabrication, wet etching, track etching## 1. INTRODUCTION

Recent progresses in nanobiotechnology attract increasing interest in nanopore fabrication. A nanopore is a small pore in an electrically insulating membrane that can be used as a molecular probe. Feasible applications include ion-pumping [1-2], DNA sequencing and separation [3-4], determination of length of polymers and separation of polymers by length [5]. The size of a nanopore (with diameter and length smaller than 20 nm) is required to be just a little larger than the molecule which is to be investigated such that a precise measurement can be obtained.

Several kinds of techniques have been developed to precisely fabricate nanopores, including ion-beam sculpting [6-7], electron beams [8], oxidation [9-10], vapour-etching [11], and track-etching [12]. However, most of these techniques usually involve costly fabrication equipments and processes. Some of them are subjected to specific material. It is not only cost ineffective but also application limited. A simpler and cheaper fabrication method is eagerly desired.

The main goal of this research is to bring up a simple and cheap method to fabricate nanopores in silicon substrate, using only a two-step etching process, including a double-side wet-etching followed by a track-etching.

## 2. TWO-STEP ETCHING

The main concept of the proposed method is a two-step etching process. The first step includes etching both sides of a silicon substrate by conventional photolithography process. After the first step, through nanopores at those pre-etched spots are fabricated using the track-etching technique and a specially designed fixture. Figure 1 and 2 schematically illustrate the procedures of the two-step etching process.

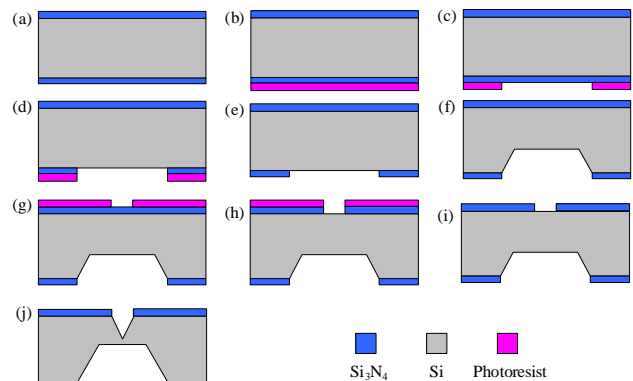

Figure 1. Schematic illustration of the double-sided etching processs

©EDA Publishing/DTIP 2007                    ISBN: 978-2-35500-000-3

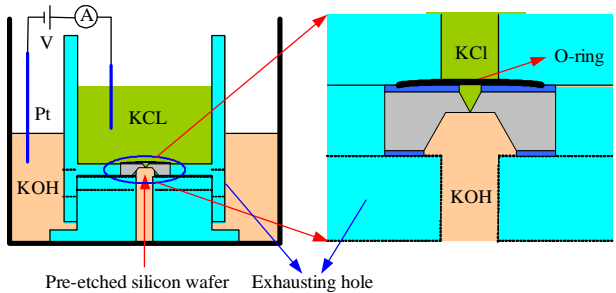

Figure 2. Schematic illustration of the track-etching setup

**(1) Double-sided wet etching**

The double-side etching as shown in Figure 1 includes thin film deposition, photolithography, inductive coupled plasma (ICP) dry-etching, and wet-etching.

**Step (a):** Depositing $Si_3N_4$ as the hard mask

The silicon wafer used is a 380 μm thick <100> p-type wafer provided by the Wafer Works Corp. The $Si_3N_4$ layers on both sides of the silicon wafer are deposited using the LPCVD process such that the residual stress is reduced and thicker films (10,000 Å) can be made. The process parameters are: temperature-850°C, pressure-180 mtorr, reaction gas- $NH_3$ and $SiH_2Cl_2$.

**Step (b) & (c):** Patterning the backside photoresist

A etch window on the backside of the $Si_3N_4$ deposited wafer is patterned by conventional photolithography process. Detail processes are listed below.

*i)* Prepare working mask for the etch window.
*ii)* Spin-coat a 7 μm thick negative photoresist (JSR THB-120N). Parameters for the spinning coating are: spinning speed of the 1st stage= 1000 rpm, spinning time for the first stage= 10 sec, spinning speed of the 2nd stage= 3500 rpm, spinning time for the 2nd stage= 25 sec (Figure 1b).
*iii)* Soft bake with temperature set at 90 °C for 5 min.
*vi)* Transfer the etch window pattern.
*v)* Hard bake at 120 °C for 7 min.

**Step (d):** Patterning the backside $Si_3N_4$ film

The ICP-RIE (Cirie-100) dry etching is adopted to transfer the etch window into the $Si_3N_4$ film. The process parameters of the ICP-RIE are: reaction gas-$CF_4$ with flow rate being 45 sccm, working pressure-5 mtorr, RF power-500 W, processing time-6 min.

**Step (e) & (f):** Processing anisotropic wet etching on the backside silicon

After removing the photoresist, the uncovered silicon surface is wet-etched by a 50% (w/w) KOH at 95°C for 4 hours. Figure 3 displays the after etching structure for a 1×1 mm² etch window.

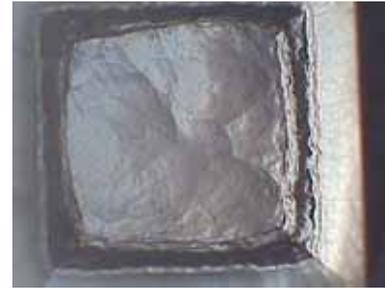

Figure 3. After etching (50% KOH at 95°C for 4 hours) structure for a 1×1 mm² etch window

**Step (g)-(j):** Repeating processes (b)-(f) on the front surface

Figure 4 is the SEM micrograph of the resulting pyramid for a 50×50 μm² etch window on the front surface. The wet-etching duration is 10.5 min.

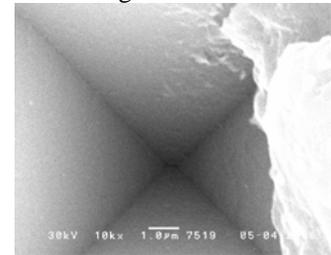

Figure 4. SEM micrograph of the pyramid for a 50×50 μm² etch window on the front surface

**(2) Track-etching**

As shown in Figure 2, a specially designed fixture is designed to hold the pre-etched silicon wafer inside it such that the front side of the wafer faces the KCL solution (1M) and the pyramid is full of KCL solution. Simultaneously, the back side of the wafer is anisotropically wet-etched by the KOH etchant (50%) which is contained by a larger glass vessel. The ionic solution KCl that has higher hydraulic pressure is used to suppress the KOH, preventing the resulting nanopore from being over etched. Since the track-etching rate is 18 nm/min, the experimental apparatus as shown in Figure 2 enables convenient process termination by only taking out the fixture as soon as the nanopore is formed.

Two Pt electrodes are put into the KOH and KCl solutions respectively and connected by a conductive wire. An electrochemical system PARSTAT 2236 (Princeton Applied Research) is employed and a 300 mV electric potential is applied to detect the current once the pre-etched cavities are etched into a through nanopore, allowing ions exchanging between the KOH and KCl solutions.

During the etching process, silicon is etched by the hydroxyl ions in the etchant to generate silicon hydroxide and free electrons. Therefore, the anode and cathode are placed in the KOH etchant and the KCl solution, respectively. The anode attracts the hydroxyl ions such that the etching rate can be slow down and a more precise nanopore can be fabricated.

In general, a nano scale pore should allow both the cation ($K^+$, 0.137nm) and anion ($Cl^-$, 0.181 nm) to penetrate simultaneously. However, due to the electric osmosis inside a nonopore, the ion diffusion induced current becomes too slight to be efficiently detected. An applied electric potential is thus needed to drive the ions traveling through the nanopore to enable real time monitoring of the nanopore formation.

## 3. EXPERIMENTAL RESULTS AND DISCUSSIONS

(1) Mask and etching duration design

The dimensions of the mask and the duration of each wet etching process need to be determined according to the wafer thickness. The thickness of the silicon wafer is 380 μm.

During the first wet-etching period, the front surface is desired to be etched to an inverted pyramid with its depth being less than the thickness of the wafer. For a 50×50 μm$^2$ etch window, the depth of the after etching pyramid is 15.988 μm. By contrast, a larger etch window that leads to an after etching trapezoid is desired; therefore, a 1×1 mm$^2$ etch window is designed. The depth of the trapezoid is controlled by the etching duration. The etching rate under a 50% KOH etchant at 95°C is 1.538 μm/min while the track-etching rate at room temperature is 18 nm/min. The thickness of the unetched silicon between the pyramid and the trapezoid should be as less as possible such that the track etching time can be minimized. In our experiment, the remaining unetched silicon is about 10.8μm in thick after going through 4 hours wet etching.

(2) Fixture design

Teflon is selected as the fixture material due to its high resistance to acid and alkali. The requirement of the Teflon fixture is that the KOH and KCl solutions can be completely sealed off, besides the pre-etched areas. Figure 5 is the blueprint of the fixture. The exhausting holes under the wafer holding place allow the residual air to be discharged.

(3) Real time monitoring of the nanopore formation processs

The remaining silicon thickness after going through the first etching process is about 10.8 μm. It takes more than 10 hours to carry out the track etching process. Figure 6 describes the real time monitoring current of the track etching process. The results indicate that a nanopre can be fabricated by the proposed two-step etching method.

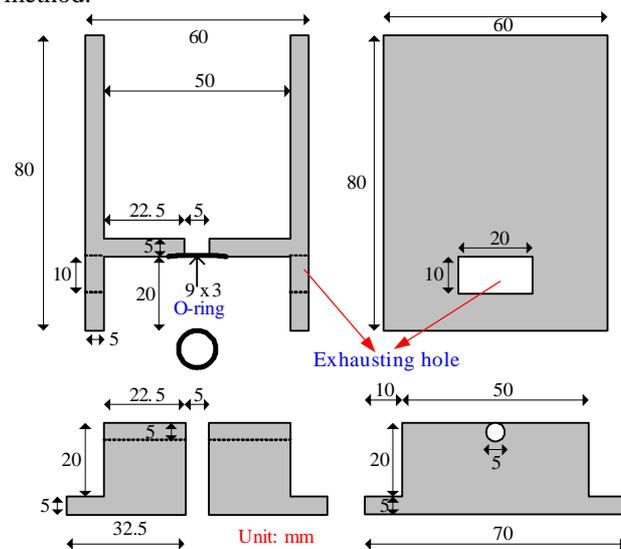

Figure 5. The blueprint of the fixture for the track etching

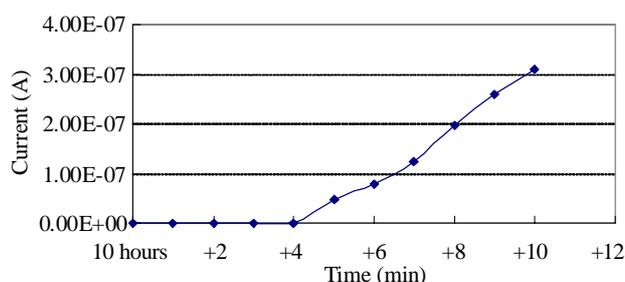

Figure 6. The real time monitoring current of the track etching process

## 4. CONCLUSIONS

In this paper, a two-step etching method to fabricate nanopores in silicon by only using the conventional wet-etching technique is presented. The two-step process includes a premier double-sided wet etching and a succeeding track-etching. A special fixture is designed to hold the pre-etched silicon wafer inside it such that the track-etching can be effectively carried out. The tracking etching is monitored in real-time by an electrochemical system that can detect the submicron scale current as soon as a nanopore is formed. Experimental results indicate that the proposed method can cost effectively fabricate nanopores in silicon.

Our future work will focus on the fabrication of a patterned nanopore array in silicon and using it to grow patterned carbon nano tube array for the field emission display applications.

**ACKNOWLEDGEMENTS**

The authors would like to thank the National Science Council of Taiwan, for financially supporting this work under Contract No. NSC-94-2212-E-005-010. The Center of Nanoscience and Nanotechnology at National Chung-Hsing University, Taiwan, is appreciated for use of its facilities.